\documentstyle[epsfig,preprint,aps]{revtex}

\begin{document}

\draft

\title{Multiple phase transitions in Ce(Rh,Ir,Co)In$_{5}$ heavy-fermion superconductors}
\author{P.G. Pagliuso, R. Movshovich, A.D. Bianchi, M. Nicklas, N.O. Moreno, J.D. Thompson, M.F. Hundley, J.L. Sarrao and Z. Fisk}
\address{Condensed Matter and Thermal Physics Group, MS K764, Los Alamos National Laboratory, Los Alamos, New Mexico 87545, U.S.A.}

\maketitle

\begin{abstract}

Magnetic susceptibility, electrical resistivity and heat capacity
data for single crystals of Ce(Rh,Ir)$_{1-x}$(Co,Ir)$_{x}$In$_{5}$
(0 $\leq$ $x$ $\leq$ 1) have allowed us to construct a detailed
phase diagram for this new family of heavy-fermion
superconductors(HFS). CeRh$_{1-x}$Ir$_{x}$In$_{5}$ displays
superconductivity(SC) ($T_{c}$ $\lesssim$ 1 K) over a wide range
of composition, which develops out of and coexists (0.30
$\lesssim$ x $\lesssim$ 0.5) with a magnetically ordered state,
with $T_{N}$ $\sim$ 4 K. For CeCo$_{1-x}$Rh$_{x}$In$_{5}$, the
superconducting state ($T_{c}$ $\sim$ 2.3 K for $x$ = 0) becomes
a magnetic state ($T_{N}$ $\sim$ 4 K, for $x$ = 1) with two phase
transitions observed for 0.40 $\lesssim$ $x$ $\lesssim$ 0.25.
CeCo$_{1-x}$Ir$_{x}$In$_{5}$ also shows two transitions for 0.30
$\lesssim$ $x$ $\lesssim$ 0.75. For those alloys in which SC is
found, a roughly linear relationship between $T_{c}$ and the
lattice parameter ratio c/a, was found, with composition as the
implicit parameter. The interplay between magnetism and SC for
CeRh$_{1-x}$(Ir,Co)$_{x}$In$_{5}$ and the possibility of two
distinct superconducting states in CeCo$_{1-x}$Ir$_{x}$In$_{5}$
are discussed.

\end{abstract}


Recently, a new series of CeMIn$_{5}$ tetragonal compounds for
M=Co, Rh or Ir have become the focus of much research due to the
discovery of heavy-fermion superconductivity in these
materials.\cite{Hegger,Petrovic,Petrovic2} In particular,
CeCoIn$_{5}$ ($\gamma$ $\approx$ 300 mJ/mole-K$^{2}$) has the
highest $T_{C}$ = 2.3 K yet observed for HFS at ambient pressure,
and the SC has been claimed as unconventional and presumably
magnetically mediated.\cite{roman} CeIrIn$_{5}$($\gamma$
$\approx$ 720 mJ/mole-K$^{2}$) also shows ambient-pressure
heavy-fermion SC at $T_{c}$ $\approx$ 0.4 K. In contrast,
CeRhIn$_{5}$ is an antiferromagnet at ambient pressure ($T_{N}$
$\approx$ 3.8 K and $\gamma$ $\approx$ 400
mJ/mol-K$^{2}$)\cite{Hegger} which evolves to a superconducting
state for P $>$ P$_{c}$ $\approx$ 16kbar and $T_{c}$ $\approx$ 2
K. As such, the CeMIn$_{5}$ family is a great opportunity to
investigate the tunability of HF ground states using external
control parameters such as pressure or doping. In this regard,
recent studies of CeRh$_{1-x}$Ir$_{x}$In$_{5}$ have revealed a
remarkably rich phase diagram which is reminiscent of the
high-$T_{c}$ cuprates.\cite{Pagliuso} The main findings of these
studies were: i)SC was observed in a very wide doping range (
0.30 $<$ x $\leq$ 1.0), ii) Magnetism ($T_{N}$ $\approx$ 3.8 K)
and SC ($T_{c}$ $\lesssim$ 1 K) $coexist$ for a significant range
of concentration (0.3 $<$ x $<$ 0.65), and iii)The maximum
$T_{c}$ was found at the boundary between magnetism and SC. Here
we report the generalization of Ref. 5 to include pseudo binary
phase diagrams for the Rh-Co and Ir-Co alloys. Our two main
results are: 1) the presence of multiple coexisting phases is a
general feature of the ground states of CeMIn$_{5}$ and 2)for
those materials that superconduct, a universal relationship
between $T_{c}$ and $c/a$, independent of alloy composition, is
found.

Figure 1 shows the generalized temperature-composition phase
diagram for Ce(Rh,Ir,Co)In$_{5}$, derived mainly from heat
capacity measurements on single crystals. SC is found in wide
ranges of $x$ for Rh-Ir, Rh-Co and Co-Ir alloys. The robustness of
the superconducting state as a function of doping for these
compounds is unusual for HFS. As discussed in Ref.5, details of
the Rh-Ir section of this phase diagram resemble that of the
high-$T_{c}$ cuprates.\cite{Pagliuso} Because CeRhIn$_{5}$ is an
antiferromagnet, and CeIrIn$_5$ and CeCoIn$_5$ are
superconductors at ambient pressure, studies of
CeRh$_{1-x}$(Ir,Co)$_{x}$In$_{5}$ allow the interplay between
magnetism and SC to be studied in detail. As Co, Rh, and Ir are
all isovalent, the parameter being tuned is not effective doping,
but rather effective magnetic coupling or hybridization. For both
Ir and Co substitutions, a comparable and large amount of doping
is required to suppress long range magnetic order. Further, both
Rh-Ir and Rh-Co alloys show ranges of concentration ( 0.3 $<$ x
$<$ 0.65 and 0.30 $\lesssim$ $x$ $\lesssim$ 0.75, respectively)
where two phase transition are observed by heat capacity
measurements. Additional resistivity and $chi_{ac}$ measurements
allowed us to identify the higher temperature transition as a
magnetic transition and the lower-T transition as SC, consistent
with the  systematic evolution of ground states of the end member
compounds(see Fig.1).

With respect to their proximity to magnetism, Ce-based HFS belong
to a class in which ordered magnetism seems to compete with SC and
in which both states have been observed concomitantly only for
small windows of parameter space.\cite{Heffner} The phase diagram
reported here is a counterexample to the above categorization
because significant ranges of coexistence of these states have
been found, and in fact, is more reminiscent of phase diagrams
associated with U-based HFS.\cite{Heffner} For
CeRh$_{1-x}$Ir$_{x}$In$_{5}$ NQR, $\mu$SR and neutrons scattering
experiments indicate $microscopic$ coexistence of magnetism and
SC and the absence of gross phase segregation.\cite{Nick}
Finally, the phase diagram of Fig.1 presents an interesting
asymmetry with respect to the interaction between magnetism and
SC: in the case of Rh-Ir, T$_c$ increases as the magnetic
boundary is approached, whereas in the Rh-Co case, T$_c$
decreases.  Thus, whether the SC in CeMIn$_5$ requires proximity
to magnetism to exist or is destroyed only by a sufficiently
strong magnetic state remains to be seen.

Turning now to CeCo$_{1-x}$Ir$_{x}$In$_{5}$, two heat capacity
anomalies are observed for 0.30 $\lesssim$ $x$ $\lesssim$ 0.60.
Zero resistivity is achieved at the high-T transition and
maintained through the low-T transition. Preliminary $\chi_{ac}$
data reveal the onset of diamagnetism  at the upper transition
and no obvious signature at the lower one. Although powder
diffraction patterns and low residual resistivity suggest
otherwise, at this point we cannot rule out the possibility of
sample inhomogeneity (such as a distribution of doping
concentration) as being the origin of the two transition. The
fact the both pure compounds CeCoIn$_{5}$ and CeIrIn$_{5}$ are
superconductors might suggest the presence of two superconducting
phase transitions.

Finally, aside from the absence of SC for compounds near
CeRhIn$_5$, one can find a quasi-linear increase of $T_{c}$ as a
function of $x$ for CeRh$_{1-x}$(Ir,Co)$_{x}$In$_{5}$ (dashed
line in Fig.1). Not coincidentally, the lattice parameter ratio
$c/a$ also increases in the sequence
CeIrIn$_{5}$-CeRhIn$_{5}$-CeCoIn$_{5}$. Figure 2 presents $T_{c}$
versus $c/a$ for Ce(Rh,Ir,Co)In$_{5}$ for those compounds in
which SC is observed. A linear dependence that spans more than a
factor of five in T$_c$ is apparent. Estimates of anisotropic
compressibility derived from thermal expansion measurements on
CeIrIn$_{5}$\cite{Gegen} allow the inclusion of $T_{c}$ for
CeRhIn$_{5}$ and CeIrIn$_{5}$ under pressure in the Fig.2. This
surprising positive linear relationship between $T_{c}$ and $c/a$
perhaps gives support for recent calculations for magnetically
mediated superconductors.\cite{Monthoux}In these calculations,
quasi-2D crystal structures and nearly commensurate
antiferromagnetic correlations are essential for producing higher
$T_{c}$s. Thus, the increasing of $c/a$ in this family of
compounds may increase the 2D character of the relevant spins
fluctuations. However, how relatively small changes in $c/a$ can
lead to a rather dramatic tuning of the ground-state properties
of these materials remains to be clarified (e.g., tuning the
f-ligand hybridization, changing the symmetry of Ce crystal-field
ground state, etc.) A linear fit to data on Fig.2 yields a slope
$\Delta$$T_{c}$/$\Delta$$c/a$ $\approx$ 70 K. Optimistically (and
naively), a member of the CeMIn$_{5}$ family with a $c/a$
$\approx$ 6 would be a room temperature superconductor.

%
%
%
%

%
%
%
%


\begin{figure}[tbh]
\centering
\includegraphics[scale=0.5]{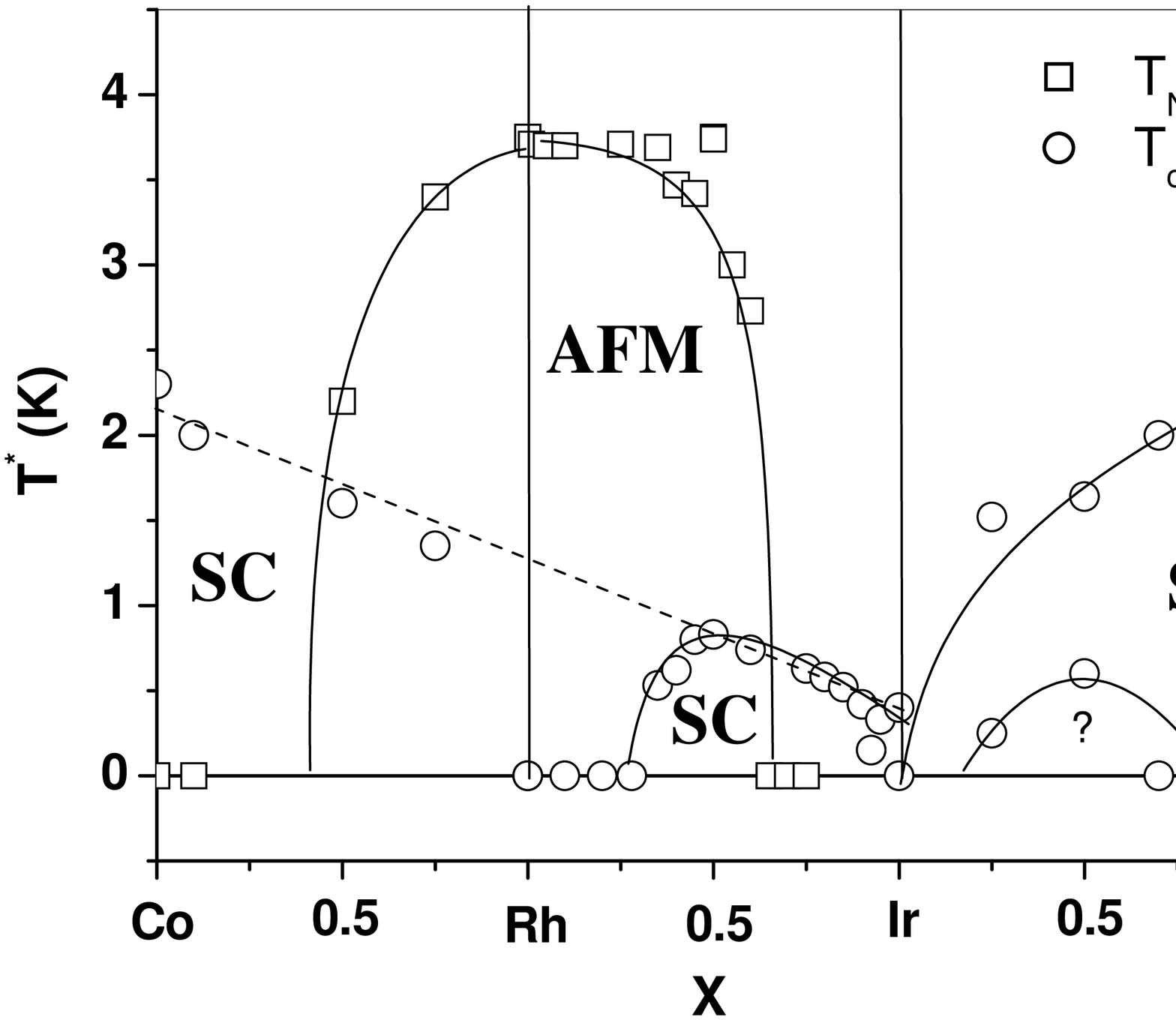}
\caption{Phase Diagram of Ce(Rh,Ir,Co)In$_{5}$}
\end{figure}

\begin{figure}[tbh]
\centering
\includegraphics[scale=0.5]{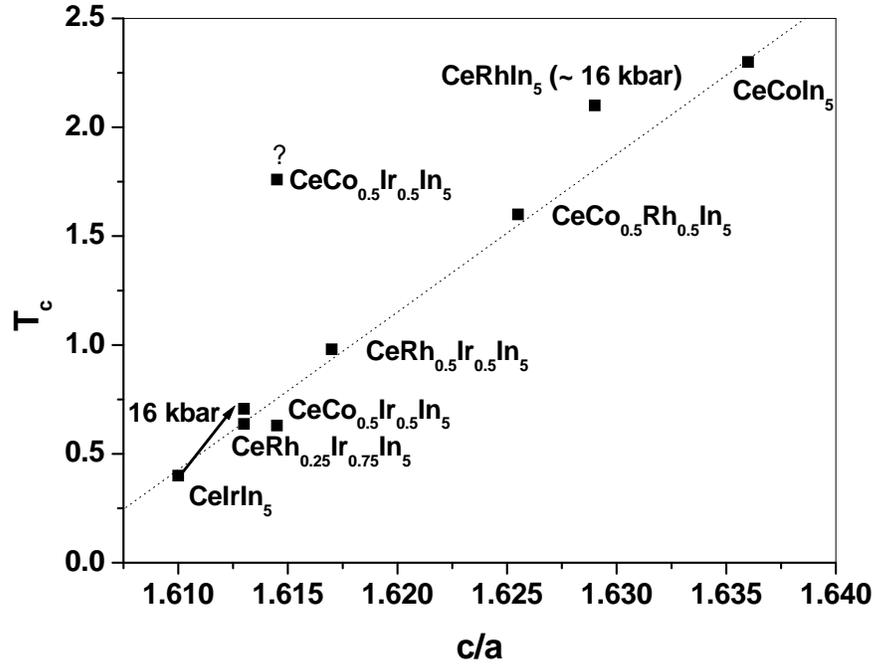}
\caption{$T_{c}$ as a function of $c/a$ for the
Ce(Rh,Ir,Co)In$_{5}$ HFS.}
\end{figure}


\begin{references}

\bibitem{Hegger} H. Hegger, et al. Phys. Rev. Lett. 84, 4986(2000).

\bibitem{Petrovic}C. Petrovic, et al. Europhys. Lett. 53 354-359 (2001).

\bibitem{Petrovic2}C. Petrovic, et al.  J. Phys.: Condens. Matter 13 L337 (2001).

\bibitem{roman}R. Movshovich et al. Phys. Rev. Lett. 86, 5152 (2001).

\bibitem{Pagliuso}P.G. Pagliuso, et al. to appear in Phys. Rev. B
Rapid. Comm.(2001).

\bibitem{Heffner}R.H. Heffner and M.R. Norman, Comm. Condens. Matter Phys. 17, 361 (1996).

\bibitem{Nick} N.J. Curro et al.; R.H. Heffner et al; W. Bao et al. unpublished.

\bibitem{Gegen}P. Gegenwart et. al. unpublished.

\bibitem{Monthoux} P. Monthoux and G.G. Lonzarich, Phys. Rev. B, 59, 14598
(1999).


\end{references}
\end{document}